\documentclass[10pt]{article}

\usepackage{setspace}
\usepackage{graphicx}
\usepackage{bm}

\newcommand{\ee}{\end{equation}}
\newcommand{\be}{\begin{equation}}
\newcommand{\la}{\langle}
\newcommand{\ra}{\rangle}
\begin{document}
\bibliographystyle{prsty}
\title{A lattice Boltzmann study of non-hydrodynamic effects in shell models of turbulence}
\author{{R. Benzi$^{1}$, L. Biferale$^{1}$\footnote{Corresponding author: L.Biferale, Dept. of Physics, University of Tor Vergata, Via della Ricerca Scientifica 1, 00133 Rome, ph +390672594595; email:biferale@roma2.infn.it},  M. Sbragaglia$^{1}$,}\\ 
{S. Succi$^{2}$ and F. Toschi$^{2}$}\\  
{\small $^{1}$ Dipartimento di Fisica and INFN, Universit\`a "Tor Vergata",}\\ 
{\small Via della Ricerca Scientifica 1, I-00133 Roma, Italy}\\
{\small $^{2}$ Istituto per le Applicazioni del Calcolo, CNR,}\\ 
{\small Viale del Policlinico 137, I-00161, Roma, Italy}}
\maketitle
\begin{abstract}
A lattice Boltzmann scheme simulating the dynamics of shell models of
turbulence is developed.  The influence of high order kinetic modes
(ghosts) on the dissipative properties of turbulence dynamics is
studied.  It is analytically found that when ghost fields relax on the
same time scale as the hydrodynamic ones, their major effect is a net
enhancement of the fluid viscosity.  The bare fluid
viscosity is recovered by letting ghost fields evolve on a much longer
time scale.  Analytical results are borne out by high-resolution
numerical simulations. These simulations indicate that the hydrodynamic
manifold is very robust towards large fluctuations of 
non hydrodynamic fields.
\end{abstract}
\maketitle
\section{Introduction}
In the recent years it has been speculated
that modern developments in (discrete) kinetic theory might yield a new
angle of attack to the problem of turbulence modeling 
\cite{RGLBE1,RGLBE2}.
The rationale behind this idea is as follows. All turbulence models
based on the Navier-Stokes hydrodynamic equations work on the
assumption of a scale separation between the resolved and unresolved
eddies \cite{BOU,LES,LESieur}.  
By resolved eddies, we mean excitations at a scale larger
than the grid size of the simulation.
Such an assumption is never satisfied in turbulent flows, 
particularly close to solid boundaries, where turbulence
production and removal are strongly unbalanced. This is the reason 
why all eddy-viscosity models fail to reproduce 
accurately turbulent statistics in strongly non-homogeneous situations.
%
Kinetic theory is at a vantage point to describe such strongly
out of equilibrium conditions, since it does not require any scale
separation between the 'fast' and 'slow' degrees of freedom.
We wish to emphasize that we are referring to
an {\it effective} kinetic theory dealing with 
the dynamics of {\it quasiparticles}, i.e. effective 
degrees of freedom of the turbulent flow \cite{BE,LBE1,BGK}.

In any eddy-viscosity model the effect of small 
eddies on the large ones results in a typical
diffusion process, only with a much enhanced turbulent diffusivity.
In contrast to eddy-viscosity, the kinetic approach is centered upon
the more general and fundamental notion of {\it relaxation},
controlled usually by a single characteristic time, $\tau$ entering in
the  Boltzmann equation. The ratio between the relaxation time and the typical
hydrodynamical time, $\tau_{h}$, is called the Knudsen number, $Kn =
\tau/\tau_h$. In the limit of small $Kn$, the Boltzmann equation converges to
the Navier-Stokes equations \cite{BE,LBGK}.
The major question is: how to derive
a suitable effective kinetic Boltzmann equation for the dynamics of 
large scale in  turbulent flows?
In principle, such an equation should be derived ab-initio 
through a renormalization-group procedure,
starting from the true Boltzmann equation for molecules. 
Preliminary attempts in this ambitious direction look very encouraging,
and yet not conclusive \cite{KARLI}.
While waiting for a rigorous derivation, a practical strategy is to
resort to discrete versions of the Boltzmann equation, now known as
Lattice Boltzmann equation (LBE), and endow it with a self-consistent
relaxation operator \cite{SCI}.
The crucial asset of LBE is that it provides a minimal form
of kinetic theory compatible with the physics of turbulent flows,
and such that it can be simulated very efficiently
on present-day computers.
This approach, sometimes called LBE-$\tau$, has been recently shown to
provide leading-edge numerical results for 
turbulent flows in real-life geometries \cite{SCI}.
Despite the impressive results, these simulations leave many theoretical
ends open, and 'whether a theory can be developed remains to be
seen' \cite{BENSCI}.
Leaving aside the important issue of numerical efficiency,
from a purely theoretical perspective, a relevant question is
whether the added value of the kinetic
approach to fluid turbulence can be linked to the
the dynamics of non-hydrodynamic fields, high order kinetic
moments of the Boltzmann distribution, sometimes called 
ghost fields \cite{BSV,VBS,DELLAR,wolf}. 
Ghost fields represent the hidden kinetic information which, although
necessary to guarantee the correct local symmetries, does not 
normally emerge to the macroscopic scale. 
In this paper we aim at investigating the role of the dynamics of
ghost fields in the kinetic approach to fluid turbulence.  
In particular, we focus our attention on 
the effects of ghost fields on the small-scale 
statistics of high Reynolds number flows. 
Unfortunately, due to limitations in computational power, it
is still impossible nowadays 
to perform numerical simulation in the fully developed
turbulent regime.  To reach high Reynolds numbers flows, with a clear
separation between the energy injection and the energy dissipation
scales, one needs to resort to some model of turbulence. A popular
class of deterministic dynamical models, widely used in recent years,
is given by ``shell-models'' \cite{SHELL}. Shell models represent the
only example where flows at high Reynolds numbers, with realistic
small-scales statistics, can be studied.  They are a good test-bed for
new theories and numerical schemes aimed at improving the
understanding of small-scales turbulent behaviour.  The paper is
organized as follows.  First we propose a Lattice Boltzmann scheme for
shell-models. Second, we study analitycally and numerically its
hydrodynamical limit with particular emphasis on the importance of
ghosts fields for the small scales dynamics.  Finally, we present
numerical and analytical results on the ``multi-relaxation'' regime,
i.e. the case when hydrodynamic fields and ghost fields have two
different relaxation properties to the local equilibrium.
\section{Shell models}
Shell models are the simplest dynamical systems featuring a realistic
picture of energy transfer from large to small scales
\cite{FRI,SHELL}.  The main advantage of shell models is that they
can be analyzed in great depth via highly accurate numerical
simulations, which permit to resolve up to six decades in momentum
space. They are non-linear deterministic models of Navier-Stokes
equations, built such as to preserve energy, helicity and phase-space
volume in the inviscid and unforced limit.  In this work we shall
focus on the following shell model \cite{sabra}:
\be\label{shelleq} 
\partial_{t}U_{n} = i k_{n}Q_{n}-\nu k^{2}_{n}U_{n}+F_{n}
\ee
where the non linear term is given by:
\be
Q_{n}=U^{*}_{n-2}U_{n-1}+\frac{b}{2}U^{*}_{n-1}U_{n+1}+\frac{(1+b)}{4}U_{n+1}U_{n+2}.
\label{sm}
\ee
In the above
$U_n$ is a complex variable representing the fluctuating velocity
field at wavenumber $k_n=2^n k_0$, and
$b$ is a free parameter fixing the physical dimension of the second inviscid invariant (here we fixed $b=-0.5$).  
In the presence of a large-scale forcing $F_n = F \delta_{n,0}$, this
model exhibits excellent scaling laws, from many aspects  indistinguishable
from those of real turbulence. 
For example,  the p-th order structure functions
\be\label{exponent}
S_{p}(n)=\la |U_{n}|^{p}\ra \sim k^{-\zeta(p)}_{n}
\ee 
are characterized by a set of scaling exponents, $\zeta(p)$, very close
to those measured on turbulent flows. Many other aspects, connected to
the velocity probability density functions, energy dissipation
statistics and multi-time multi-scale correlation functions are also
in good agreement with what measured in experimental and numerical
studies of Navier-Stokes equations. For these reasons shell models
have represented a unique occasion to investigate small-scale
turbulent statistics without the difficulties of the original
Navier-Stokes equations.  Despite their apparent simplicity, a full
systematic analytical control of the small-scale statistical behaviour
is still lacking.  Recently, a series of promising closure attempts
based on stochastic closures have been proposed \cite{ben1,ben2,ben3}.
Yet, they cannot be considered conclusive. The importance of Lattice
Boltzmann schemes for the shell-model  (\ref{sm}) is therefore
 twofold.  First, they may shed some lights on the complex
multi-time dynamics of the hydrodynamical limit, second they may be
useful to control, and optimize, convergence to the hydrodynamic
limit, which may be useful  also for LB schemes of Navier-Stokes
equations.
\section{LBE shell model}
In this section we shall develop a discrete kinetic model whose
hydrodynamic limit is precisely the shell model eq.(\ref{sm}).  To this
purpose we introduce a 5-speed  lattice Boltzmann scheme in the wave number space, $k_n$,
obeying the following dynamic equations:
\be\label{BGK}
\partial_{t} f_j(k_{n}) + i k_n f_j(k_{n})= -\frac{1}{\tau} (f_j(k_{n})-f^{eq}_j(k_{n}))
\ee
where $f_j=[f_0,f_{1},f_2,f_{3},f_4]$ is the discrete distribution
associated to the discrete speeds
$c_j=[0,1,-1,2,-2]$.
The local equilibrium is given by
\be
f_{j}^{eq}(k_{n})=w_{j}[R_n+c_j U_n+({c_j}^{2}-1)D_n]
\ee
where $w_j=[6/12,2/12,2/12,1/12,1/12]$ are normalized weights.
The equivalent of the macroscopic density and
 velocity fields are defined as follows:
\be\label{densita}
R_n=\sum_{j=0}^{4} f_j(k_{n}) \hspace{.5cm} U_n=\sum_{j=0}^{4} f_j(k_{n}) c_j.
\ee
The third macroscopic field, the analogue of the traceless momentum flux
tensor, must be adjusted in such a way as to reproduce the non-linear term 
$Q_n$ in the shell model. After simple algebra, one derives:
\be
2D_n+R_n=Q_n.
\ee 
Let us notice that the shell-model equations we want to mimick 
are written in terms of a complex variable, $U_n$. 
Therefore, here the $f_j(k_{n})$'s loose the nature of probability
density functions they usually have in LB schemes in real space
(complex distribution functions have been already used in the past
to simulate quantum mechanics \cite{PHYSD,OUP}). 

The first consequence is that in order to keep the 
``macroscopic density'' for shell $n$,  $R_n$, 
constant in time, one has to modify the streaming term in (\ref{BGK})
for rest particles as follows: 
\be
\partial_{t} f_{0}(k_{n}) = -ik_{n}(2f_{3}(k_{n})-2f_{4}(k_{n})+f_{1}(k_{n})-f_{2}(k_{n}))-\frac{1}{\tau}(f_{0}(k_{n})-f^{eq}_{0}(k_{n}))
\ee
By a linear transformation, we move to the momenta representation
for the stress tensor, $S_n=\sum_j f_j(k_{n}) c_j^2$, and 
two ghost fields, $A_n=f_4(k_{n})-f_3(k_{n})$ and $B_n=f_3(k_{n})+f_4(k_{n})$.
The resulting  equations are:
\be\label{R}
\partial_{t}R_{n}=0
\ee
\be\label{U}
\partial_{t}U_{n}=ik_{n}S_{n} 
\ee
\be\label{S}
\partial_{t}S_{n}=ik_{n}U_{n}+6ik_{n}A_{n}-\frac{1}{\tau}(S_{n}-Q_{n})
\ee
\be\label{A}
\partial_{t}A_{n}=2ik_{n}B_{n}-\frac{1}{\tau}(A_{n}-\frac{1}{3}U_{n})
\ee
\be\label{B}
\partial_{t}B_{n}=2ik_{n}A_{n}-\frac{1}{\tau}(B_{n}-\frac{1}{4}Q_{n})
\ee
The set of equations (\ref{R}-\ref{B}) is our kinetic shell model.
From the first three equations, we obtain 
\be\label{madre}
\partial_{t}U_{n} + \tau \partial_{tt} U_n = 
ik_{n}Q_{n}-\tau k^{2}_{n}U_{n}-6 \tau k^{2}_{n}A_{n}.
\ee
This is the master equation of our treatment.
First of all, we inspect its hydrodynamic limit.
To this purpose, we notice that in the limit 
$\tau \rightarrow 0$ the ghost field $A_{n}$ collapses onto
its attractor, the velocity field:
\be
\label{U3}
\lim_{\tau \rightarrow 0} A_{n}=\frac{U_{n}}{3} 
\ee
so that (\ref{madre})  delivers:
\be
\partial_{t} U_{n}=i k_{n} Q_{n}-3 \tau k^{2}_{n}U_{n}-\tau \partial_{tt}U_{n}.
\ee
It is therefore seen that, upon neglecting the term 
$\tau \partial_{tt} U_{n}$, which is indeed higher order in the Knudsen
number, the correct shell model is reproduced in the limit
$Kn \rightarrow 0$, with the ghost field contributing a factor
$2 \tau$ to the flow viscosity.
A perturbative expansion in $\tau$ of all fields 
appearing in (\ref{madre}), reveals that the
effect of ghost fields at second order in $Kn$, yields a 
non-conservative contribution of the form
\be
\tau^{2} k^{3}_{n} Q_{n}.
\ee
The finite-Knudsen regime is therefore characterized by the interplay
of {\it three} distinct terms, namely:
\be
k_{n}Q_{n} \hspace{.3in} (inertial \; term)  
\ee
\be
\tau k^{2}_{n}U_{n} \hspace{.3in} (dissipative\;  term)
\ee
\be
\tau^{2}k^{3}_{n}Q_{n} \hspace{.3in}  (ghost\;  contribution). 
\ee
Dimensional matching of these competing terms delivers 
the relevant crossover scales in Fourier space:
\begin{itemize}
\item Dissipative scale (Dissipation=Inertia):
\be
k_{d} \sim \frac{1}{\tau^{3/4}}
\ee
\item Ghost scale (Ghost term=Inertia):
\be
k_{g} \sim \frac{1}{\tau}.
\ee
\end{itemize}
These relations show that $k_g \sim \frac{k_g}{\tau^{1/4}} >> k_d$ for
$\tau << 1$ which means that the ghost fields cannot play any role on
the dissipation properties of the system since they do not reach up to
the dissipative scale.  In order to elicit a non-trivial role for the
ghost fields, we need to realize the condition $k_g < k_d$.  This
necessarily leads to a generalized LBE in which ghost fields relax on
their own timescale, longer than the hydrodynamic one.  The minimal
such choice is to define {\it two} relaxation times: $\tau_{\nu}$ and
$\tau_g$, for hydrodynamic and ghost fields respectively.  Since we aim
at a fully turbulent regime, we shall consider for $\tau_{\nu}$ the
smallest possible values compatible with grid resolution.  The
relaxation time for the ghost field will then be changed in order to
investigate its effects on the dissipation properties of the system.
\section{Multi-relaxation shell BGK model}
The simplest multi-relaxation kinetic shell model takes the following
two-time form:
\be\label{R2}
\partial_{t}R_{n}=0
\ee
\be\label{U2}
\partial_{t}U_{n}=ik_{n}S_{n}
\ee
\be\label{S2}
\partial_{t}S_{n}=ik_{n}U_{n}+6ik_{n}A_{n}-\frac{1}{\tau_{\nu}}(S_{n}-Q_{n})
\ee
\be\label{A2}
\partial_{t}A_{n}=2ik_{n}B_{n}-\frac{1}{\tau_{g}}(A_{n}-\frac{1}{3}U_{n})
\ee
\be\label{B2}
\partial_{t}B_{n}=2ik_{n}A_{n}-\frac{1}{\tau_{g}}(B_{n}-\frac{1}{4}Q_{n}).
\ee
This set of equations is easily reproduced 
by going back to earliest LB formulations, in 
which collisional effects were taken into account 
through a scattering matrix $M_{ji}$
describing the interaction between the $j$-th and $i$-th populations:
\be\label{GBGK}
\partial_{t} f_j(k_{n}) +i k_{n} f_j(k_{n})= M_{ji} (f_i(k_{n})-f^{eq}_i(k_{n})).
\ee
Following the top-down procedure introduced in \cite{HSB}, we can
construct a scattering matrix with eigenvalues
$\lambda=\lbrace 0,0,-1/\tau_{\nu},-1/\tau_g,-1/\tau_g \rbrace$ 
and a corresponding set of kinetic eigenvectors, $V_j^{(k)}$, 
$k=0,4$, associated with the set of fields 
$R_{n},U_{n},S_{n},A_{n},B_{n}$, respectively:
\[
f_j(k_{n})= R_{n} V_j^{(0)} + U_{n} V_j^{(1)} + S_{n} V_j^{(2)} + A_{n} V_j^{(3)} + B_{n} V_j^{(4)}.
\]
This corresponds to a partition of the five-dimensional kinetic
space into a hierarchy of two conserved quantities 
($R_{n},U_{n}$), one quasi-conserved
(transport) quantity ($S_{n}$) and two ghost fields ($A_{n},B_{n}$).
Using (\ref{U2}) and (\ref{S2}) we obtain:
\be\label{madre2}
\partial_{t}U_{n} = ik_{n}Q_{n}-\tau_{\nu} k^{2}_{n}U_{n} - 6 \tau_{\nu} k^{2}_{n}A_{n} -\tau_{\nu} \partial_{tt} U_{n}.
\ee
where the dependence on $\tau_{g}$ is implicitly hidden 
within the fields $A_{n}$ and $U_{n}$. 
It is worth pointing out that at this stage
we are still dealing with {\it exact} equations.
In order to get a first guess on the dynamics of ghost fields in this case,
we make the approximation of imposing
steady-state conditions on the ghost fields equations 
(\ref{A2},\ref{B2}):
\be\label{steady1}
0=2ik_{n}B_{n}-\frac{1}{\tau_{g}}(A_{n}-\frac{1}{3}U_{n})
\ee
\be\label{steady2}
0=2ik_{n}A_{n}-\frac{1}{\tau_{g}}(B_{n}-\frac{1}{4}Q_{n}).
\ee
This yields:
\be
\label{pezzoghost}
A_{n}=\frac{U_n}{3P(k_{n}\tau_{g})}+\frac{ik_n \tau_g Q_{n}}{2P(k_{n}\tau_{g})}
,\hspace{.3cm} P(k_{n}\tau_{g})=1+4k^{2}_{n}\tau^{2}_{g}.
\ee
This steady-state approximation must be understood as an estimate for the
mean value of the fields. We show later, by direct numerical simulations
of the eqs. (\ref{R2}-\ref{B2}), that the prediction extracted 
out of (\ref{steady1}) and (\ref{steady2}), yields the 
correct qualitative and quantitative statistical behaviours. 
It is also quickly checked that in the limit $\tau_g = \tau_{\nu} << 1$ the
relation (\ref{pezzoghost}) reduces to the expression (\ref{U3}), as it should.
For further analysis it proves convenient to 
explore the behaviour of (\ref{pezzoghost}) in the
two regimes of small and large scales separately.
\subsection{Large scales}
Setting $\tau_g=1$, large scales are identified by the condition
$k_{n}<<1$. In this regime, the denominator
of (\ref{pezzoghost}) simplifies,  $P(k_{n}\tau_{g}) \rightarrow 1$, and
we obtain: 
\be
-6\tau_{\nu}k^{2}_{n}A_{n} \approx -3 i k^{3}_{n} \tau_{g} \tau_{\nu} Q_{n}-2 \tau_{\nu} k^{2}_{n} U_{n}
\ee
Upon substituting this in the master equation
(\ref{madre2}) we get:
\be\label{grandescala}
\partial_{t}U_{n} = ik_{n}Q_{n}(1-3 i k^{2}_{n} \tau_{g} \tau_{\nu}) 
-3 \tau_{\nu} k^{2}_{n}U_{n}- \tau_{\nu} \partial_{tt} U_{n}.
\ee
It is therefore apparent that, since $\tau_{\nu}<< \tau_g=1$ 
and $k_{n}<<1$, the relative correction to the 
convective term, $3 \tau_g \tau_{\nu} k_n^2$, 
is  negligible.
As a result, we come to the conclusion that
the large-scale regime is virtually uncontaminated by the ghost fields.

\subsection{Small scales}

Small scales are identified by the condition
$k_{n}>>1$, again with the position $\tau_{g}=1$. 
In this regime the denominator $P(k_{n}\tau_{g})$ is dominated by
the $k_n^2$ term, so that the master equation delivers:
\be\label{pezzoghostsmall}
-6\tau_{\nu}k^{2}_{n}A_{n} \approx -3 i k_{n}\frac{\tau_{\nu}}{\tau_{g}}Q_{n}-\frac{2 \tau_{\nu}}{\tau^{2}_{g}} U_{n}
\ee
Upon substituting in (\ref{madre2}), we obtain
\be
\partial_{t}U_{n} = ik_{n}(1-\frac{\tau_{\nu}}{\tau_{g}})Q_{n}
-\tau_{\nu} k_n^2 U_n -\frac{2 \tau_{\nu}}{\tau^{2}_{g}} U_{n}-\tau \partial_{tt} U_{n}.
\ee
>From this expression we notice that the renormalized convective term is
still conservative, with a renormalized factor
$1-\tau_{\nu}/\tau_g \sim 1$. 
The viscous term becomes $-\tau_{\nu} k^{2}_{n}U_{n}$, corresponding to a viscosity
$\nu = \tau_{\nu}$, i.e. three times lower than in the previous case.
Apart from the usual second order time derivative of the velocity field, the
remaining piece, $-{2 \tau_{\nu}/\tau^{2}_{g}} U_{n} $, 
is a sub-leading damping term,
due to the combined effect of high wavenumbers and
long ghost relaxation time. In other words, ghosts can act at scales larger
than the dissipative scale, but their amplitude is suppressed
by a factor $\tau_{\nu}/\tau_g$ and consequently
they gently disappear from the scene, leaving the system with the 
bare ghost-free viscosity $\tau_{\nu}$.
Summarizing, the present analysis leads to the following
predictions:
\begin{itemize}
\item Hydrodynamic scenario: $\tau_{g}=\tau_{\nu}<<1$, 
ghost fields are enslaved to their local equilibrium values.
They contribute an extra term $2 \tau_{\nu}$ to the fluid viscosity
and do not affect the convective terms to any appreciable extent.
At small but finite Knudsen numbers they are 
confined to sub-dissipative scales only and 
cannot produce any further appreciable effect.  
As a result, the correct hydrodynamic limit is recovered, with
an enhanced viscosity $\nu = 3 \tau_{\nu}$.

\item Non-hydrodynamic scenario: $\tau_{\nu} \rightarrow 0,\tau_{g}=1$, 
ghost fields are no longer enslaved to the fluid velocity. 
They receive contributions from the velocity field and the
non-linear term $Q_{n}$ at {\it all} 
scales through the propagator $P(k_{n} \tau_g)$.
As a result, they exhibit high-frequency, small-amplitude 
fluctuations, which do not affect the large scale behavior
of the system because they are suppressed by a $\tau_{\nu}/\tau_g$ factor. 
The correct hydrodynamic limit is still recovered,
with a bare viscosity three times smaller than in the the previous case,
 $\nu_0= \tau_{\nu}$.
\end{itemize}
\section{Numerical results}
The theoretical scenario discussed in the previous section has been
tested against numerical simulations of the kinetic shell model.
As a first test, we have simulated the kinetic shell model in the
hydrodynamic regime, namely $\tau_{\nu}=\tau_g=5 \times 10^{-4}$, with 
$k_n=2^{n-13}$, $n=1,25$.
For the sake of a quantitative comparison, the same simulations have been
repeated with the original shell model (\ref{shelleq})
 (with a viscosity $\nu=3 \tau_{\nu}$).
In Figure \ref{FIG1} we show the energy spectra for the two cases.
An excellent agreement between the LB and the shell model simulations
is observed across the {\it whole} range of scales, except for scales
well inside the dissipative range.  In the inset of
figure (\ref{FIG1}) we also present a check of the enslaving relation
(\ref{U3}) by plotting the ratio between the ghost field and the
velocity field, $3Re(A_n)/Re(U_n)$ for two typical wavenumbers, at
large scales, $n=6$ and at small scales close to the dissipative
cut-off, $n=18$.  Notice how the relation (\ref{U3}) is perfectly
verified at large scales, while some, small, deviation from slaving is
observed at the end of the inertial range.  This confirms our theoretical
analysis, namely that ghost fields are completely enslaved to their
equilibrium values and do not affect the hydrodynamic behaviour of the
turbulent system. Only strongly dissipative physics is affected by the
ghost dynamics. The global stability is not changed.

More interesting, is to explore numerically the non-hydrodynamic
regime, $\tau_g \sim O(1)$.  We have performed a set of simulations
with, $\tau_{\nu}=5 \times 10^{-4}$, $\tau_g \sim O(1)$.  The
corresponding spectra for the velocity field are shown in Figure
\ref{FIG2}, where the case $\tau_{\nu}=\tau_g$ is also reported to
highlight the effect of the reduced viscosity from $\nu=3 \tau_{\nu}$
to $\nu_{0}=\tau_{\nu}$.  As predicted by our theoretical analysis,
the LB model in the non-hydrodynamic regime reproduces turbulent shell
dynamics with the correct hydrodynamic viscosity $\nu_{0}= \tau_{\nu}$ while
the ghost field tends to decrease in intensity by going to smaller and
smaller scales. Now ghosts are decoupled from the velocity
fluctuations and the enslaving relation (\ref{U3}) is no longer
satisfied. Still, they do not have any significant impact on the
dynamics of the velocity field because their intensity is
negligible (see inset of Figure
\ref{FIG2}).\\  A deeper insight into the role of ghost fields at all
scales, inertial and dissipative, can be gained by inspecting the
structure functions (\ref{exponent}).  In particular, in figure
\ref{FIG3} we show the fourth and 
sixth order flatness $F^{(4)}_{n}= \frac{\la U^{4}_{n}
\ra}{\la U^{2}_{n} \ra^{2}}$ and 
$F^{(6)}_{n}= \frac{\la U^{6}_{n} \ra}{\la U^{2}_{n} \ra^{3}}$ as a
function of $n$ for the hydrodynamic and non-hydrodynamic regime. The
shell-model results are also presented.  First we observe that both
quantities have a clear dependency on the scale, that is a sign of
intermittency in the velocity statistics. 
Second, all numerical results agree in
the two regimes, the only difference being an increased
inertial range extension when ghosts fields decouple.

\section{Conclusions}

Summarizing, we have presented a detailed analysis of the effects of
non-hydrodynamic (ghost) fields on the statistical properties
of hydrodynamic turbulence, within the framework of shell models.
As a first result, we have shown that if ghost fields relax on the
same time-scale as the hydrodynamic one, $\tau_g = \tau_{\nu}$ 
(the common scenario in current real-space LB simulations), then 
in the hydrodynamic limit where
this scale is sent to zero, the ghost fields contribute to  the
hydrodynamic viscosity with a ratio $2:1$ with respect
to the weight of the hydrodynamic fields: $\nu = 2 \tau_g + \tau_{\nu}$. 
Higher order ghost contributions, at finite relaxation times, 
are segregated to sub-dissipative scales, so that no further
effects on the hydrodynamic behaviour can result.
The non-hydrodynamic 'overtake' scenario ($k_g<k_d$) 
can be realized by allowing ghost fields a longer 
lifetime than hydrodynamic modes.
It is then found that ghost fields do {\it not} spoil
the inertial physics (large scales) up to a negligible correction 
$\tau_{\nu}/\tau_g$. The dissipative properties of the turbulent system are
also reproduced, but now without the ghost contribution, leading 
to an hydrodynamic bare
viscosity smaller than in the previous case: $\nu_0 = \tau_{\nu}$. 

Our analysis supports the counter-intuitive notion that letting
ghosts 'alive' on long time scales leads to a reduction of 
the viscosity, as compared to the case
in which ghosts are frozen to their hydrodynamic equilibria.


Once extrapolated to the framework of real-space lattice Boltzmann
models, our findings provide a motivation towards multirelaxation
models, as opposed to the currently popular single-time relaxation BGK
model. This come-back of multirelaxation models has been invoked by
other groups as well \cite{LUO}, based on motivations of improved
(linear) stability.

It is tempting to speculate that real-space LB simulations
of two and three-dimensional turbulence might profit by 
moving (back) to a multi-relaxation scenario
in which the hydrodynamic scale is kept to its minimum fixed by grid
resolution, while the ghost fields time-scale is made much longer.

On a similar vein, one may speculate that making the hydrodynamic
and ghost time scales respond self-consistently to turbulence observables,
but with distinct functional dependencies, may prove beneficial also
for LBGK-${\tau}$ simulations with turbulence modeling \cite{SCI}. 
As a final remark, our results provide a clear evidence 
that the hydrodynamic manifold is very robust against 
large fluctuations of non-hyrodynamic fields.

\section{Acknowledgments}
Illuminating discussions with S. Ansumali, H. Chen, I. Karlin, S. Orszag 
and V. Yakhot are kindly acknowledged.

\begin{figure}[t!]
\begin{center}
\includegraphics[draft=false,scale=1.0]{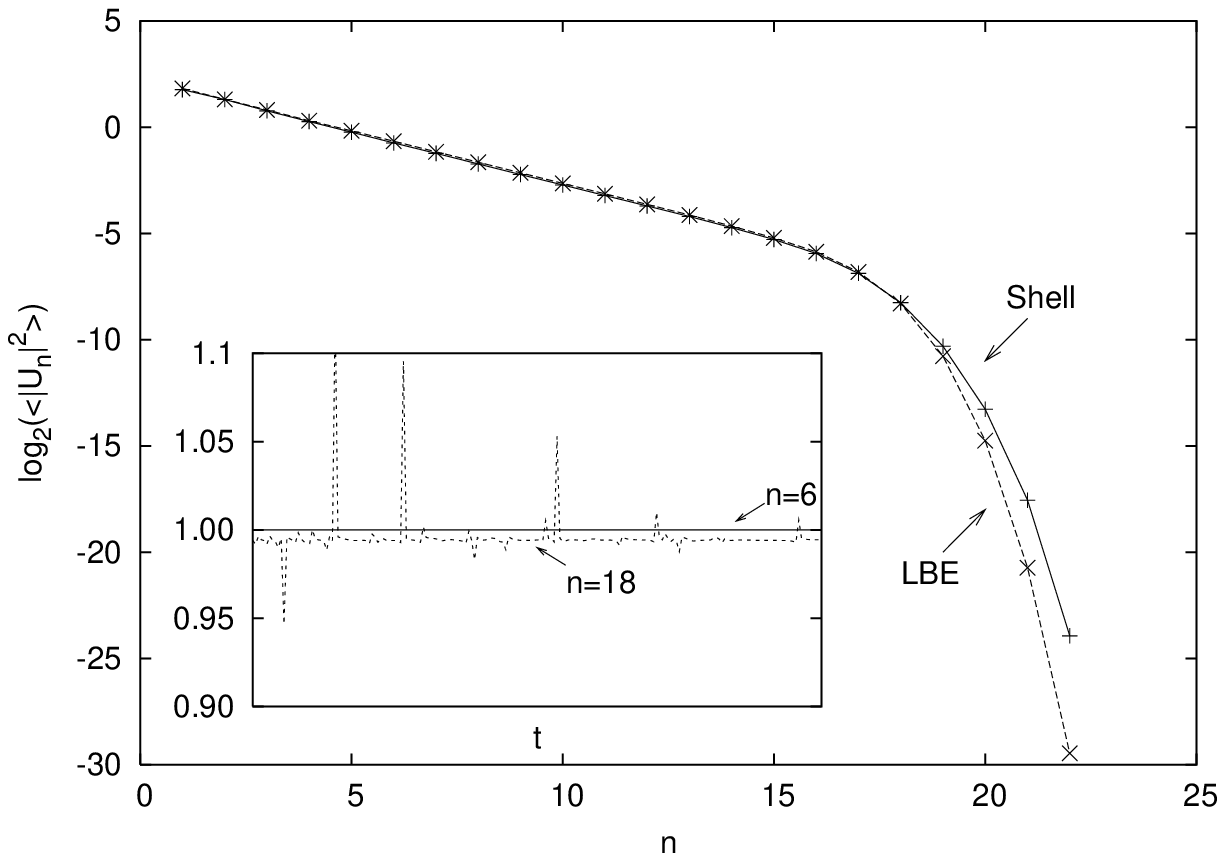}
\end{center}
\caption{}
\label{FIG1}
\end{figure}

\begin{figure}[t!]
\begin{center}
\includegraphics[draft=false,scale=1.0]{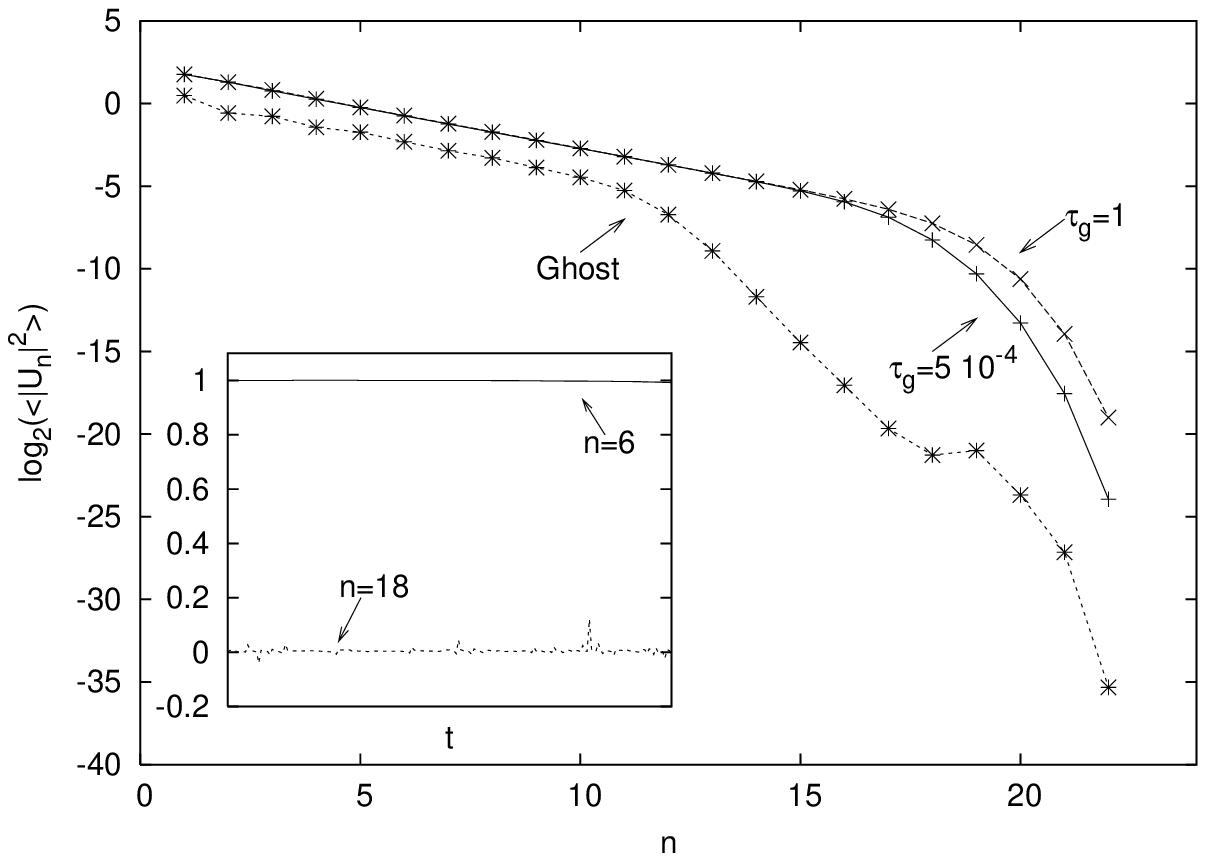}
\end{center}
\caption{}
\label{FIG2}
\end{figure}

\begin{figure}[t!]
\begin{center}
\includegraphics[draft=false,scale=1.0]{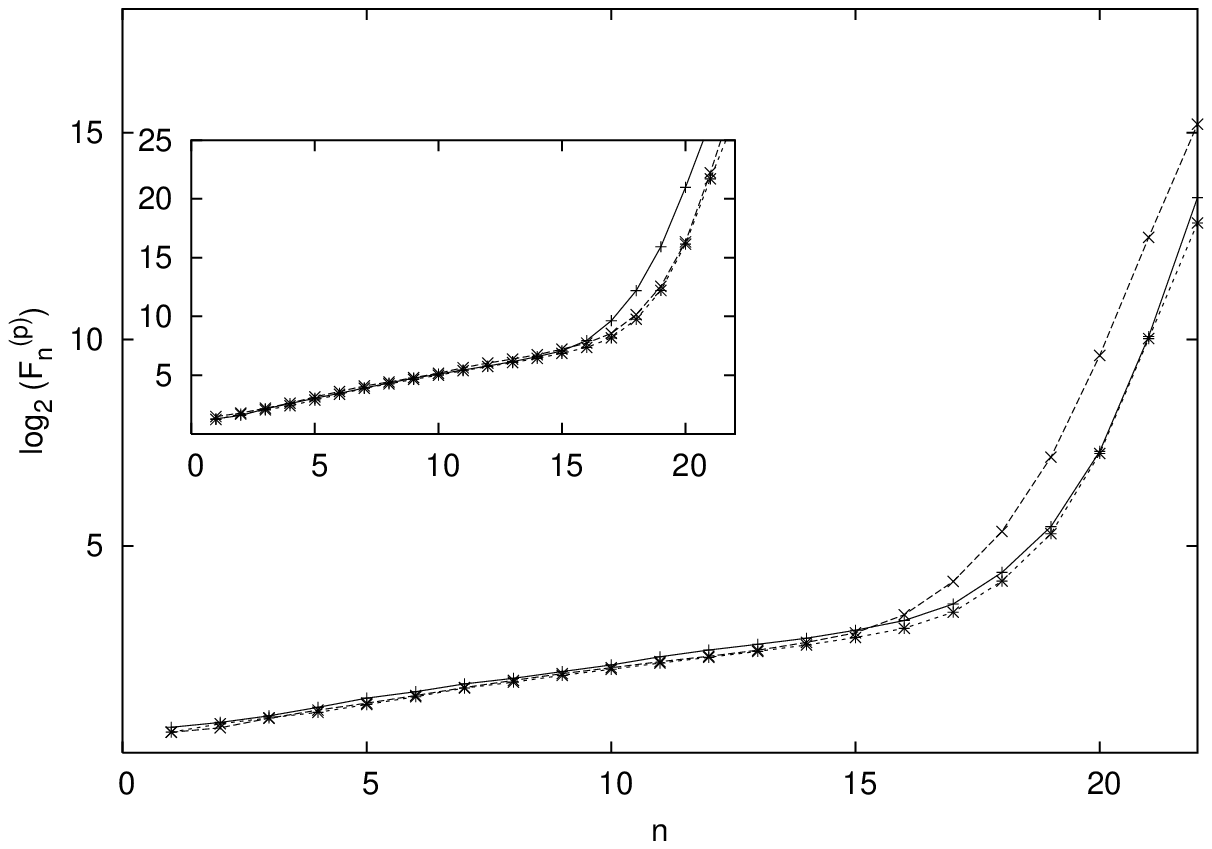}
\end{center}
\caption{}
\label{FIG3}
\end{figure}

\newpage \centerline{FIGURE 1} Comparison between the original shell
model and our LB model. We plot $\log_{2}\la |U_{n}|^{2} \ra$ vs $n$
for the shell model with viscosity $\nu=3 \tau_{\nu}$ ($+$) and the LB
kinetic model with $\tau_{\nu}=\tau_g=5 \times 10^{-4}$ ($\times$).
Both models have the same forcing acting on the first two shells.
Inset: check of the ghost-velocity slaving in the hydrodynamic regime.
We plot $Re(3 A_{n}(t))/Re(U_{n}(t))$ vs $t$ for shell index $n=6$
(straight line) and $n=18$ (dotted line) with $\tau_{\nu}=\tau_g=5
\times
10^{-4}$. Notice the small deviation observed at the smallest scale.\\
\centerline{FIGURE 2} Velocity spectra for hydrodynamic and
non-hydrodynamic LB regime. We plot $\log_{2} \la |U_{n}|^{2} \ra$ vs
$n$ for the LB model with the following choice of parameters:
$\tau_{\nu}=\tau_g=5 \times 10^{-4}$ ($+$) and $\tau_{\nu}=5 \times
10^{-4}$, $\tau_g=1$ ($\times$). Notice the increase of the
inertial-range extension in the multi-relaxation LB model because the
ghost field ($\star$) is completely negligible at small scales. Inset:
check of the ghost-velocity slaving in the non-hydrodynamic regime. We
plot $Re(3 A_{n}(t))/Re(U_{n}(t))$ vs $t$ for shell index $n=6$
(straight line) and $n=18$ (dotted line) with $\tau_g=1$. Notice now,
at difference from the case of fig.~\ref{FIG1}, that the velocity and
the ghost fields at small scales are decoupled, being the overall
intensity of the ghost field
negligible.\\
\centerline{FIGURE 3} Analysis of the flatness $F^{(p)}_{n}={\la
  U^{p}_{n} \ra}/{\la U^{2}_{n} \ra ^{p/2}}$ for different values of
$p$.  We plot $\log_2 F^{(4)}_{n}$ vs $n$ in our kinetic model for the
following choice of parameters: $\tau_{\nu}=\tau_g=5 \times 10^{-4}$
($\times$) and $\tau_{\nu}=5 \times 10^{-4}$, $\tau_g=1$ ($+$); to
emphasize the correct behaviour of the system we also plot the same
quantity in the case of the original shell model with viscosity
$\nu_{0}=\tau_{\nu}$ ($\star$). Inset: the same cases but for the
sixth order flatness $F^{(6)}_{n}={\la U^{6}_{n} \ra}/{\la U^{2}_{n}
  \ra^{3}}$.  \newpage

\newpage

\newpage

\begin{thebibliography}{99}
  
\bibitem{RGLBE1} H. Chen, S. Succi, S. Orszag, {\it Phys. Rev. E.}
  {\bf 59}, R2527, (1999).
  
\bibitem{RGLBE2} S. Succi, O. Filippova, H. Chen, S. Orszag, {\it J.
    Stat. Phys} {\bf 107 }, 261, (2002).
  
\bibitem{PHYSA} S. Succi, I. Karlin, H. Chen, S. Orszag, {\it Physica
    A} {\bf 280}, 92, (2000).

\bibitem{BOU} C. Meneveau and J. Katz, {\it Annu. Rev. Fluid Mech.}
  {\bf 32}, 1, (2000).
  
\bibitem{LES} Geurts, M. (2004): {\em Elements of direct and
    large-eddy simulation}, R.T. Edwards Inc..
  
\bibitem{LESieur} Lesieur, M. and Metais, O. (1994): Ann. Rev. Fluid
  Mech. {\bf 28}, 45--82.
  
\bibitem{KARLI} S. Ansumali, I.V. Karlin, and S. Succi, {\em Kinetic
    theory of turbulence modeling: Smallness parameter, scaling and
    microscopic derivation of Smagorinsky model}, LANL
  cond-mat/0310618, (2003), to appear in Physica A.
  
\bibitem{SCI} H. Chen, S. Kandasamy, S. Orszag, R. Shock, S. Succi,
  and V. Yakhot,
  Science {\bf 301}, 633--636, (2003).
  
\bibitem{BENSCI} R. Benzi, {\em Getting a grip on turbulence}, Science
  {\bf 301}, 605-606, (2003).
  
\bibitem{BSV} R. Benzi, S. Succi, M. Vergassola, {\it Phys. Rep.} {\bf
    222}, 145, (1992);
  
\bibitem{VBS} M. Vergassola, R. Benzi, S. Succi, {\it Europhys. Lett.}
  {\bf 13}, 411, (1990);
  
\bibitem{DELLAR} P. Dellar, {\it Phys. Rev. E} {\bf 65}, 036309,
  (2002);
  
\bibitem{wolf} D. A. Wolf-Gladrow, ``Lattice-Gas Cellular Automata and
  Lattice Boltzmann Models'', Lecture Notes in Mathematics, 1725,
  Springer (Berlin), 2000.
  
\bibitem{SHELL} L. Biferale, {\it Ann. Rev. Fluid Mech.} {\bf 35} 441,
  (2003).
  
\bibitem{FRI} U. Frisch, {\it Turbulence, The legacy of A.
    Kolmogorov}, Cambridge Univ. Press, (1996).
  
\bibitem{sabra} V. S. L'Vov, E. Podivilov, A. Pomyalov, I. Procaccia,
  D. Vandembroucq, {\it Phys. Rev.} E {\bf 58} 1811, (1996).
  
\bibitem {ben1} R. Benzi, L. Biferale, F. Toschi, {\it J. Stat. Phys.}
  {\bf 113} 738, (2003).
  
\bibitem {ben2} R. Benzi, L. Biferale, M. Sbragaglia, F. Toschi, {\it
    Phys. Rev.} E {\bf 68} 046304, (2003).
  
\bibitem {ben3} R. Benzi, L. Biferale, M. Sbragaglia, {\it J. Stat.
    Phys.} {\bf 114} 137, (2004).
  
\bibitem{BE} C. Cercignani, {\it Theory and application of the
    Boltzmann equation}, {\it Elsevier}, New York, (1975).

\bibitem{LBE1} G. McNamara, G. Zanetti, {\it Phys. Rev. Lett.} {\bf
    61}, 2332, (1988);

\bibitem{BGK} P. Bhatnagar, E. Gross, M. Krook,
{\it Phys. Rev. A} {\bf 94}, 511, (1954).

\bibitem{LBGK} Y. Qian, D. d'Humieres, P. Lallemand,
{\it Europhys. Lett.} {\bf 17}, 479, (1992).




\bibitem{PHYSD} S. Succi, R. Benzi,
{\it Physica D} {\bf 69}, 327, (1993).

\bibitem{OUP} S. Succi,
{\it The Lattice Boltzmann Equation for Fluid Dynamics and Beyond},
{Oxford University Press}, (2001);

\bibitem{HSB} F. Higuera, S. Succi, R. Benzi,
{\it Europhys. Lett.} {\bf 9}, 345, (1989);

\bibitem{LUO} P. Lallemand, L.S. Luo, 
{\it Phys. Rev. E} {\bf 61}, 6546, (2000);



\end{thebibliography}
\end{document}